\documentclass{article}

\usepackage{microtype}
\usepackage{graphicx}
\usepackage{subfigure}
\usepackage{booktabs} 
\usepackage{multirow}
\usepackage{amsmath}
\usepackage{amssymb}

\usepackage[dvipsnames]{xcolor}
\definecolor{mygray}{gray}{0.6}
\definecolor{caribbeangreen}{rgb}{0.0, 0.8, 0.6}
\usepackage{hyperref}

\usepackage[accepted]{icml2021}

\def\Jian#1{\textcolor{red}{Jian: #1}}

\icmltitlerunning{Non-Autoregressive Electron  Redistribution Modeling for Reaction Prediction}

\begin{document}

\twocolumn[
\icmltitle{Non-Autoregressive Electron  Redistribution Modeling for Reaction Prediction}



\icmlsetsymbol{equal}{*}

\begin{icmlauthorlist}
\icmlauthor{Hangrui Bi}{equal,pku}
\icmlauthor{Hengyi Wang}{equal,pku}
\icmlauthor{Chence Shi}{mila,udem}
\icmlauthor{Connor Coley}{mit}
\icmlauthor{Jian Tang}{mila,cifar,hec}
\icmlauthor{Hongyu Guo}{nrc}
\end{icmlauthorlist}

\icmlaffiliation{pku}{Peking University}

\icmlaffiliation{mit}{MIT}
\icmlaffiliation{nrc}{National Research Council Canada}

\icmlaffiliation{mila}{Mila - Quebec AI Institute}
\icmlaffiliation{udem}{University of Montr\'eal}
\icmlaffiliation{hec}{HEC Montr\'eal}
\icmlaffiliation{cifar}{CIFAR AI Research Chair}

\icmlcorrespondingauthor{Jian Tang}{ jian.tang@hec.ca}
\icmlcorrespondingauthor{Hongyu Guo}{hongyu.guo@nrc-cnrc.gc.ca}

\icmlkeywords{Machine Learning, ICML}

\vskip 0.3in
]



\printAffiliationsAndNotice{\icmlEqualContribution}

\begin{abstract}

Reliably predicting the products of chemical reactions presents a fundamental challenge in synthetic chemistry. 
Existing machine learning approaches typically produce a reaction product by sequentially forming its  subparts or intermediate  molecules. Such autoregressive  methods, however,   
not only require a pre-defined order for the incremental construction but preclude
the use of parallel decoding for efficient computation.
To address these issues, we devise a  non-autoregressive  learning paradigm that predicts reaction in one shot. 
Leveraging the fact that chemical reactions can be described as a redistribution of electrons in molecules, we formulate a reaction as an arbitrary electron flow  
 and predict it with a novel  multi-pointer decoding  network. 
Experiments on the USPTO-MIT dataset show that,  our approach has established a new state-of-the-art 
top-1 accuracy  and  achieves 
  at least 27 times inference speedup  over  the state-of-the-art methods. 
 Also, our predictions are  easier for chemists to interpret owing to predicting the electron flows. 
\end{abstract}

\section{Introduction}

 
     Reaction prediction~\cite{corey1969computer}, which aims to predict  the resulting chemical outcomes from  given reactants and reagents, is  a fundamental problem in computational chemistry. Reliably predicting such outcomes 
     enables chemists to analyze the feasibility of chemical reactions and design optimal synthesis routes for target molecules, which is of crucial   importance for synthesis planning, drug discovery, and  material invention. Nevertheless, reaction prediction has remained a foundational challenge owing to the fact that  a typical reaction may involve nearly 100 atoms~\cite{WLN2017jin}, which makes fully exploring all possible transformations  intractable. 
      \begin{figure}[ht]
        \centering
        \includegraphics[width = 0.48\textwidth]{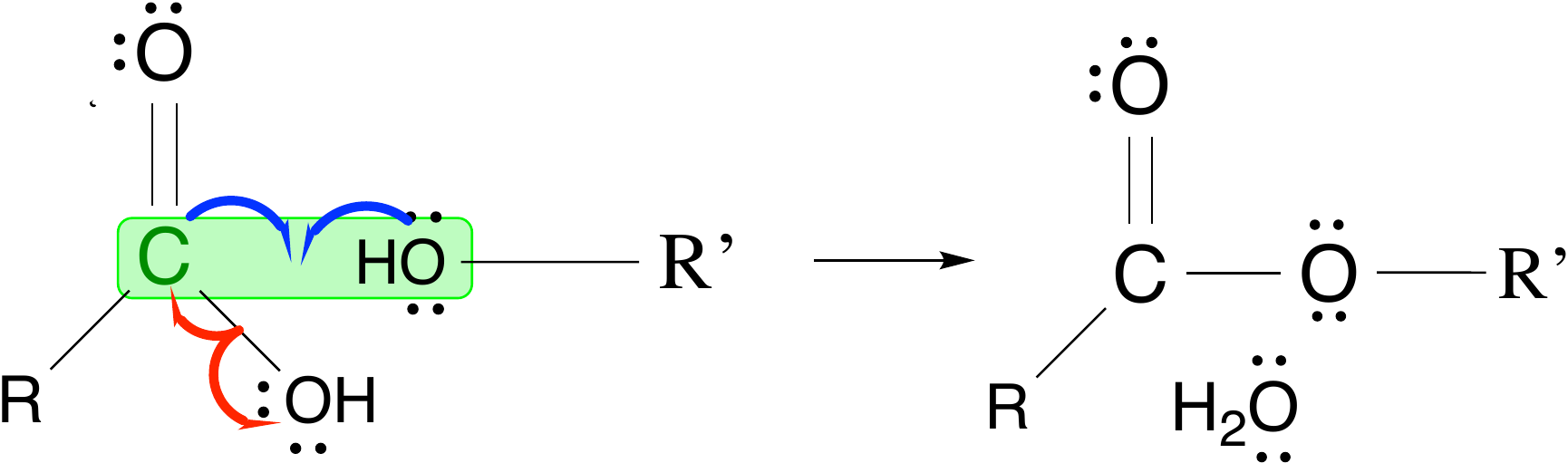}
        \caption{
        Illustration\footnotemark
         of the electron movements 
        in a reaction,  where lines represent single covalent bonds (each comprising 2 shared electrons) and a pair of   dots 
        representing a lone pair. 
        Each pair of curly arrows show the  movement of one electron pair, 
        which results in the {\em bond broken} (red curly arrow) and {\em bond formed} (blue curly arrow) in the reaction.        Such electron redistribution yields  the formation and breaking of chemical bonds that change the reactant (left) into product  (right). Our approach here aims to model all the  movements of electron pairs (i.e., curly arrays)  in one shot. 
        }
        \label{electronflow}
    \end{figure}
         
         \vspace{-2mm}

         \footnotetext{This  formalism is inspired by, but does not accurately represent, a true arrow pushing mechanism; 
        Hydrogen is  neglected for the simplicity of our model as  that of  other graph-based approaches owing to that  the  Hydrogen can be  inferred with valences and electrons provided.}
    Recently, profound successes have   been achieved by applying machine learning  to cope with the  aforementioned challenge in reaction prediction~\cite{WLN2017jin,struble2020CASP,shi20d}.
    For example, \cite{schwaller2018found,schwaller2019moltransformer} formulate reaction prediction as a problem of sequence translation by representing molecules as  SMILES strings, and Transformer based architecture is used for the translation. More advanced methods~\cite{coley2019wldn5,do2019gtpn,sacha2020megan} represent molecules as graphs and formulate reaction prediction as iterative graph transformation process. 
    Generally, these methods  generate subparts or intermediate molecules of a reaction product in a sequential fashion, through leveraging  successive decoding   steps. 

         Despite their dramatic  successes,  these  state-of-the-art end-to-end  methods have a major shortcoming, thus hindering their wider applications. That is, these  strategies embrace  an autoregressive decoding procedure, which produces  reaction subparts or intermediate molecule graphs  sequentially, each conditioning on previously generated ones. 
         Although autoregressive models can model the dependency between the sequential subparts,  there is no unambiguous  principle way of  linearizing  the sequence of steps for constructing a molecular  graph, and one   failed step in such successive  procedure could invalidate the entire synthesis outcomes. 
         Furthermore, 
         such iterative generations  hinder   parallel decoding for efficient computation. 


    To cope with the aforementioned  limitations, we propose a novel framework for reaction prediction, which predicts reaction outcomes  in one shot. 
    We leverage the fact that chemical reactions can be described as a redistribution of electrons in molecules. As illustrated in Figure~\ref{electronflow}, such electron movement results in the formation and breaking of chemical bonds that convert the reactants into product molecules~\cite{articleherge}. We here aim to model the \emph{simultaneous}  electron flows in reactants, as such 
    predicting the reaction product is  a byproduct of capturing the  electron rearrangement in the reactants. 
        In a nutshell, we first formulate an edge in a molecular graph as a pair of shared electrons (i.e., a covalent bond) and bond transformations as electron flows. We then propose a novel electron flow principle  
        to describe arbitrary and parallel electron flows in molecules. We further  implement this principle based on the conditional variational autoencoder framework~\cite{SohnLY15} with a novel 
         multi-pointer decoding network to model the incoming and outgoing electron movement probabilities for each atom.  This results in   modeling   arbitrary,  non-linear electron flows and  simultaneous graph transformations for reaction prediction, hence forming reaction products in one shot. Also, our method   possesses beneficial  interpretability through showing how  the reactants react via the reactivities of electrons.
    We refer to this model as Non-autoregressive Electron Redistribution Framework (NERF).   

    
    We evaluate our model using  the benchmarking USPTO-MIT dataset.  Our empirical studies show that 
    our approach has established  a new state-of-the-art top-1 accuracy. Moreover, our method achieves    at least 27  times faster for inference  over the state-of-the-art approaches. 
     Our experiments also indicate that the  latent variables introduced in our method enable the generation of  diverse multi-modal outputs, resulting in top-k accuracy comparable to its autoregressive counterparts.
     We also demonstrate that 
      due to the prediction of   electron flows, 
      our reaction predictions are   easier for chemists to interpret.

\begin{figure*}[t]
        \centering
        \includegraphics[width = 1\linewidth]{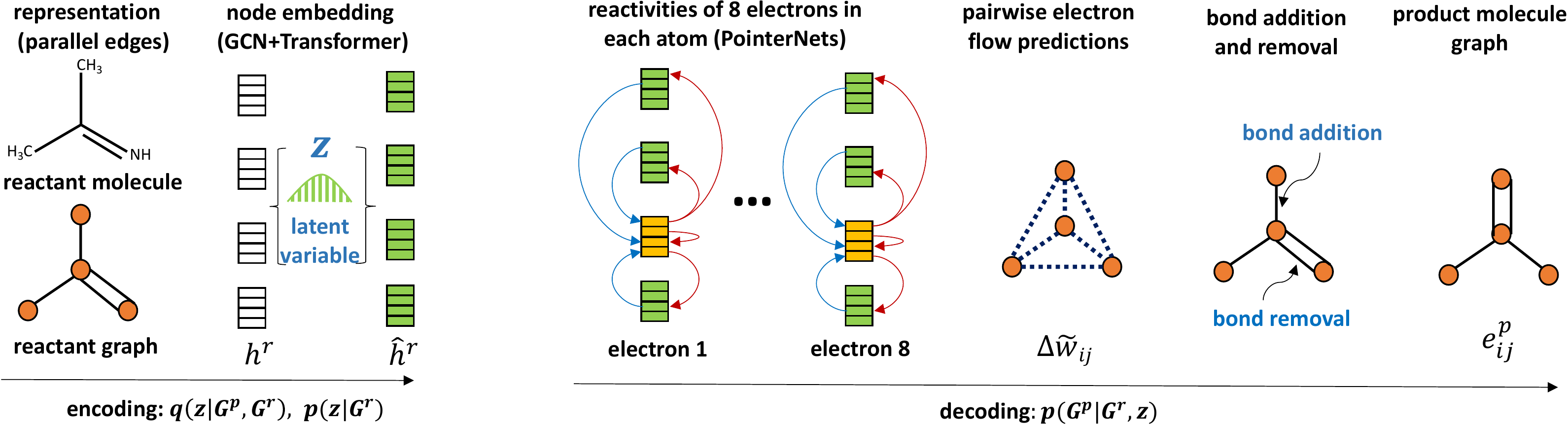}
       \caption{Framework of NERF. The left-bottom depicts the parallel edges in our graph representation $G^{r}$. Encoder  is on the left: the node embeddings are generated by a conditional variational auto-encoder CVAE~\cite{SohnLY15}. Decoder is on the right: the node embeddings   leverage a set of directional PointerNets~\cite{vinyals15pointer} to use attention weights to estimate  the flow of electron pairs between all the atom pairs in the reaction simultaneously. These electron flows thus reflect the changes of edges and then can   be converted  into the addition and removal of edges in the reactant molecule to form the  product molecule $G^{p}$ (right). }
        \label{framework}
\end{figure*}	

\section{Related Work}




Data-driven reaction prediction approaches have historically relied on reaction templates, which define sub-graph matching rules for  similar organic reactions. 
Products are generated by first mapping  reactants to a set of  templates and then applying the pre-defined transformations   encoded in the selected  templates~\cite{ segler2017modelling,segler2017neural,wei2016neural}, optionally followed by a re-ranking step \cite{coley2017prediction}.
Despite their successful application to synthesis planning, these methods suffer from poor generalization on unseen molecular structures owing to the use of rigidly-defined templates to describe how different chemical substructures may react. 

To overcome the limitation of template-based approaches, 
various 
template-free methods~\cite{WLN2017jin,kayala2011reaction,bradshaw2018electro,WLN2017jin,coley2019wldn5,schwaller2018found,schwaller2019moltransformer,qian2020integrating} have recently been introduced. 
For example,  
two-stage learning methods have been introduced to achieve promising results~\cite{WLN2017jin,qian2020integrating}. 
One downside of these approaches is that they   involve multi-stage learning and  cannot be optimized end-to-end.  
To this end, most state-of-the-art strategies typically embrace an end-to-end,  template-free  learning paradigm. For example,  
\citet{schwaller2018found,schwaller2019moltransformer}  leverage  a sequence-to-sequence framework 
to translate  SMILES~\cite{weininger1988smiles} representations of reactant graphs to those  of product graphs. 
\cite{bradshaw2018electro,do2019gtpn, sacha2020megan} formulated reaction prediction as a sequence of graph transformations on  molecule graph. 
Such autoregressive methods, however, not only require a pre-defined order for the incremental construction but also hinder parallel decoding for efficient computation. 
Our strategy overcomes these challenges by embracing a non-autoregressive  learning paradigm  to generate reaction products in one shot, enabling end-to-end training with parallel decoding for efficient computation.  

To the best of our knowledge, our approach is the first attempt to  modeling electron flows for non-autoregressive chemical reaction prediction. 
The only work we are aware of on predicting molecule electron flow  is  ~\cite{bradshaw2018electro}. 
 Nevertheless, their approach is limited to a subclass of chemical reactions with  ``linear'' electron flows in an autoregressive format. 
  In contrast, our strategy predicts electron redistribution in one shot. It embraces  an arbitrary electron flow principle to capture the parallel electron movement, thus the simultaneous bond making and breaking, in reactant molecules.



\section{The Non-Autoregressive Electron  Redistribution  Framework} 
In this paper, we formulate the chemical reaction prediction task as a simultaneous  electron redistribution problem. We solve this problem by predicting electron flows around each atom in parallel  with a conditional variational autoencoder.  
Specifically,
we first employ a graph neural network to encode the reactant graphs, where multiple covalent bonds  between  atom pairs are represented by parallel binary  edges --each  depicts two electrons that are shared between the atom pair. In the decoder, the goal is to model the incoming and outgoing electron movement probabilities for each atom (i.e., graph node) in the reactants. Finally, the electron reactivities of the reactants are converted into the addition  and removal of edges in the reactant graphs, forming the reaction products. The proposed framework is depicted in Figure~\ref{framework}, and will be discussed in detail next. 

\vspace{-2mm}

\subsection{Notation and Problem Formulation}
\label{notations}

\noindent\textbf{Molecule Graph with Parallel Edges}
We represent 
a set of organic molecules 
as  an undirected graph $G = (V, E, f)$, where $V$ and $E$ denote the set of heavy atoms (excluding hydrogen) and bonds, respectively. 
  Each node $v_i \in V$ is associated with a fixed-length feature vector ${f_i}$, each dimension of which corresponds to a certain property of the atom $v_i$ such as atom type or formal electric charge. 
 Unlike previous work where different bond orders are represented by one edge with different features or labels,  
our method  treats different bonds as multiple {\em parallel binary edges}, e.g., a double bond is represented as two parallel binary edges  (see far left of  Figure~\ref{framework}).  

\noindent\textbf{Chemical Reaction as Electron Redistribution}
In this paper, a chemical reaction is described as  a pair of molecular graphs 
$(G^{r}, G^{p})$, where $G^{r} = (V, E^r, f^r)$ is the set of 
reactants and $G^{p} = (V, E^{p}, f^{p})$  the products we aim to predict. 
Note that, these two share the same set of nodes $V$. 
From the chemistry perspective, chemical reactions can be defined as a redistribution of electrons in  molecules~\cite{ci00017a011}. In other words, the forming of $E^{p}$  can be described as the rearrangement of electrons in $E^{r}$, which  reflects the addition or removal of edges in $E^{r}$ and hence the forming of the products $E^{p}$. 

\noindent \textbf{Edge  Change  as Electron Flow in Molecule} 
From a chemistry perspective, each covalent single bond in organic (carbon-based) molecules  represents two electrons that are shared between the atom pair that the bond connects. See illustration in  Figure~\ref{electronflow}. In this paper, we denote $e_{ij}$ ($\geq 0 $ and $  \in E$) as the number of binary edges (i.e., shared electron pairs)  between nodes $v_i$ and $v_j$, ($v_i \in V,  v_j \in V$), i.e. the number of electrons of atom $i$ that are shared with atom $j$. To this end, we allow self-loops in $G$ and let $e_{ii}$ denote the number of 
{\em lone pairs} in atom $i$  (illustrated as a    dot pair around an atom in Figure~\ref{electronflow}), namely valence electrons that are not shared with other atoms. 

From this perspective,  an edge change in a molecule is caused by a valence electron's reactivity. It can form a new bond by sharing with another atom (or form a lone pair through coupling with a sibling atom of its own), or  break an existing bond by withdrawing the sharing.  Also, based on  a chemical rule of thumb known as the Octet Rule, 
valence electrons that are active in reactions and form bonds is typically no larger than 8. This motivates our aim of 
 modeling the incoming and outgoing electron movement probabilities of each of the 8 valence electrons in each atom during reaction. Consequently,  such electron flows  reflect the change of edge number before and after the reaction for each  atom pair ($i$, $j$),  ($v_{i}, v_{j} \in V$), namely
\begin{equation}
\label{deltae}
    \Delta e_{ij} = e_{ij}^p - e_{ij}^r.  
\end{equation} 
In other words, through capturing the reactivity probabilities of the  8 valence electrons in each atom,  we can  approximating $\Delta e_{ij}$ directly. By doing this, our method can simultaneously access all the edge changes in the reactant molecule, thus produce the reaction products in one shot. Such   non-autoreggressive nature 
  distinguishes our strategy from previous work in ~\cite{bradshaw2018electro},  which can only handle 
  a subclass of chemical reactions that have  linear electron flows only. 

\subsection{The Conditional Variational AutoEncoder Framework}

Our goal is to model the probability of product graphs $G^p$ conditioning on the reactant graphs $G^r$, i.e., $P(G^p|G^r)$. To model the uncertainty of reaction, we introduce a latent variable $z$ and base the whole model on the conditional variational auto-encoder framework~\cite{SohnLY15}. For training, we aim to maximize the log-likelihood of each training example, i.e., $\log P(G^p|G^r)$. In practice, it is usually computational intractable to maximize the likelihood $\log P(G^p|G^r)$, and a common approach is to maximize an evidence-lower bound (ELBO) of the log-likelihood as below:
\begin{equation}
\label{cvaeequ}
\begin{aligned}
    \log p(G^p|G^r) \geq {\mathrm{E}}_{q(z|G^p,G^r)} [\log p(G^p|G^r, z)] -\\
    KL(q(z|G^p,G^r)||p(z|G^r)), 
\end{aligned}
\end{equation}
where $q(z|G^p,G^r)$ is an encoder which defines the variational distribution of the true posterior distribution $p(z|G^p,G^r)$, $p(z|G^r)$ is the prior distribution of the latent variable $z$, and $p(G^p|G^r,z)$ is the decoder. To optimize the lower-bound, we can use the reparametrization trick introduced in the variational auto-encoder~\cite{kingma13vae}. Next, we will introduce our encoder and decoder respectively.  

\subsubsection{Encoder}
\label{encodingGCN}

As each reactant is represented as a graph, a natural choice is to encode the atoms of the whole reactants with graph neural networks~\cite{schlichtkrull2018rgcn,rong2020grover}, which have been widely studied to learn representations of different kinds of graphs. Let $k$ $\in \mathbb{N}$ be the embedding dimension and $h^{r(l)}_{i} \in \mathbb{R}^{k\times 1}$  the node embeddings of the $i^{th}$ atom  at the $l^{th}$ layer computed by the GCN $(h^{r(0)}_{i} = f_{i})$. 

At the $l^{th}$ layer, each node $i$ first aggregates the messages from  
its  neighboring  nodes, namely $n  \in \mathcal{N}_i$:  
\begin{equation}
    m_{i}^{r(l)} = \text {RELU}(W^{(l)} \cdot \text{SUM}\{h_{u}^{r(l-1)} \vert u \in \mathcal{N}_i\}), 
\end{equation}
where $ W ^{(l)}\in \mathbb{R}^{k\times k}$ is the trainable weight matrix. The aggregated messages are further combined with the node representation itself by simply taking the summation, yielding a new node representation:
\begin{equation}
    h_{i}^{r(l)} = m_{i}^{r(l)} + h_{i}^{r(l-1)}. 
\end{equation}

As graph convolutional networks only capture the local dependency between atoms, to further model the long-range dependency between atoms, we further apply a Transformer Encoder~\cite{vaswani2017transformer} on top of the atom representations learned by GCNs, similar to  \cite{rong2020grover}: 
\begin{equation}
h^{r} = \text{Transformer-Encoder}(h^{r(L)}), 
\label{seminodeembedding}
\end{equation}
where $h^{r(L)}$ are the atom representations in the final layer of the GCN, and $h^{r}$ are the final atom representations of reactants. Similarly, we encode the product graphs with the same neural encoder, obtaining the atom embeddings $h^{p}$ for each atom in $G^{p}$.

Once we have the representations for the reactant and product graphs $h^r$ and $h^p$, we then  use them to define our variational distribution $q(z|G^p, G^r)$. 
Specifically, we first pass the $h^{r}$ and $h^{p}$ through a Transformer-Decoder  to compute the cross attention between $h^{r}$ and $h^{p}$. This Transformer-Decoder takes $h^{p}$ as the  input sequence of the first decoder layer   and $h^{r}$ as the last output  sequence of the encoder. 
Essentially, the atom representations of product graphs $h^p$ are treated as memory, and the atom representations of reactants $h^r$ are treated as queries, which are updated by attending to $h^p$.
After that, a mean pooling is further applied to all the atom representations of the reactants, yielding a global representation $h^z$ of the reaction: 
\begin{equation}
\begin{aligned}
    h^{z}=\text{MEAN(Transformer-Decoder}(h^r,h^p)). 
    \end{aligned}
\end{equation}
Next, a fully connected layer with  ReLU activation
is further applied to $h^z$ to define the mean and variance of the latent variable: 
\begin{equation}
\label{kldiv}
\begin{aligned}
    &\mu  = W_{\mu}\text{ReLU}(h^z)+b_{\mu},\\
    &\text{log}\sigma = W_{\sigma}\text{ReLU}(h^z)+b_{\sigma}.\\
\end{aligned}
\end{equation}
where $ W_{\mu}, W_{\sigma}\in \mathbb{R}^{k\times k}$ and $b_{\mu}, b_{\sigma} \in \mathbb{R}^{k\times 1}$ are trainable parameters.
Therefore, the variational distribution is defined as $q(z|G^p, G^r)=\text{Normal}(\mu, \sigma)$. For the prior distribution $p(z|G^r)$, we simply assume it is a simple standard  Gaussian distribution, which does not depend on $G^r$.

\subsubsection{Decoder}
\label{decodingsec}

Next, we introduce how to decode the product graph $G^p$ based on the latent variable $z$ and the reactant graphs $G^r$, i.e., $p(G^p|G^r, z)$. As introduced previously, since the atoms are exactly the same between $G^r$ and $G^p$, we need to generate the edge difference between the two sets of graphs. We first generate a new set of atom representations for the reactant graphs (denoted as $\hat{h}^r$) by taking the summation of each atom representation $h_i^r$ and the latent variable $z$, followed by a global transformation through a  Transformer encoder layer:  
\begin{equation}
\label{sampleH}
\hat h^r_{i} = h^r_{i} + z,
\end{equation}
\begin{equation}
 \hat h^r= \text{Transformer-Encoder}(\hat h^r).
\end{equation}
     Next, these embeddings are  then 
     used to  generate the edge numbers $ e_{ij}^{p}$ for each atom pair ($i$, $j$) in the products. Recall from Section~\ref{notations} that, in order to predict the product graphs, we simply need to model each atom's electrons activities. According to the Octet Rule as discussed in ~\ref{notations},  the number of active valence electrons in an atom is typically at most 8, and
the reactivity of each electron  includes either {\em forming} a  new bond or {\em breaking} an existing bond with another atom (including the atom itself).  
     We here attain this goal  by
      modeling the bond formation and breaking   probabilities associating to each electron in   each atom using  a set of PointerNets~\cite{vinyals15pointer}. 
     
PointerNet is proposed for selecting an item from a set of items using attention mechanism.  In our setting, for each electron of an atom $i$ in reactants $G^r$,   PointerNet can compute the attentions between atom $i$ and all the  atoms  in $G^r$ (including $i$). The attentions are calculated through a softmax function, thus each of those attention weights can be interpreted as a probability. That is, through the attention weights, a PointerNet provides us the   probabilities of atom $i$ initiating an  electron flow  with all the atoms in the graph. 


Consequently, for each electron $d$,  we deploy two PointNets: one capturing its probability (denoted $w^{+d}$)   of  bond formation and  another the probability (denoted $w^{-d}$) of  bond breaking.  We denote these two types of PointerNet as    {\em BondFormation}  and {\em BondBeaking}  PointerNets, respectively. That is, in total our model has 8  BondFormation  PointerNets and 8  BondBeaking   PointerNets, and all the 16 PointerNets are shared among all the atoms in the reactants. 
 Formally,  
give a set of nodes $\hat h^r$ in the reactants $G^{r}$, a  PointerNet  
can compute an attention weight to reflect the probability of an electron flow   from  atom $i$ to any  atom $j$ in $G^r$:  
        \begin{equation}
w_{ij}^{(+d)} = \text{PointerNet}_d^+( \hat h^r, i, j), d\in{1,2,\cdots, 8}
    \end{equation}
    \begin{equation}
w_{ij}^{(-d)} = \text{PointerNet}_d^-( \hat h^r, i, j), d\in{1,2,\cdots, 8}
    \end{equation}
    
where $w_{ij}^{(+d)}, w_{ij}^{(-d)} \in [0,1]$.

    With the attentions  computed by the 16 PointerNets,  the overall 
    electron reactivity for an atom pair ($i$,  $j$) 
    then can be  approximated by  calculating the difference between the summation of  the  8   BondFormation  attentions  and  that of the 8  BondBreaking  attentions: 
    \begin{equation}
    \label{ppointer}
        \begin{aligned}
        &\Delta  \tilde w_{ij} = \sum_{d=1}^8 w^{(+d)}_{ij} - \sum_{d=1}^8 w^{(-d)}_{ij}. \\
        \end{aligned}
    \end{equation}
    Here, the first and second summations represent the additions and the removals of edges, respectively. 
    Doing so,  $\Delta \tilde w_{ij}$ thus reflects the  change of edge number between the atom pair ($i$, $j$). 
 Next, 
    we then add it to the  existing number of edges, i.e., $e^r_{ij}$, resulting in  the  new edge number $\hat e_{ij}$ between the atom pair ($i$,$j$): 
    \begin{equation}
    \label{wwij}
        \hat e_{ij} = e^r_{ij} + \Delta \tilde w_{ij}.
    \end{equation}

Finally, we define our likelihood function as:
\begin{equation}
\begin{aligned}
   \log p(G^p|G^r,z) &= \sum_{(i,j)\in E^p} \log p(e_{ij}^p|G^r,z)\\
   &\propto -\sum_{(i,j)\in E^p} (e_{ij}^p-\hat{e}_{ij})^2.
    \end{aligned}
\end{equation}
    
\textbf{Generation} 
In practice, given a set of reactant graphs $G^r$, to generate the product graphs, $\hat e_{ij}$ is   rounded to the closest integer to obtain  the  edge number between the atom pair, namely $e^p_{ij}$ :  
    \begin{equation}
    \label{edgeGeneration}
        e_{ij}^p = \text{round}(\text{min}(\hat e_{ij}, \hat e_{ji})).
    \end{equation}
Doing so,  we then have the resulting modified  graph $ G^{p}$. 
During evaluation, a reaction is correctly predicted if the ground truth product is a subgraph of our prediction $G^{p}$ (more  detailed illustration will be presented in Figure~\ref{explainable} in the Experiment section).

\subsection{Atom Feature 
Construction and
Prediction
}
We assign each graph node (i.e., atom) with a unique identifier as its positional embedding~\cite{you2019positionaware}. This ensures that atoms in different chemical environments have distinct node representations.
The input features of atoms include atom type, charge, aromaticity, segment embedding and positional embedding. There is no edge feature in our model.

Our proposed NERF framework can  easily  incorporate domain knowledge owing to the fact that each atom has its own embedding $h^{r}$.  
In addition to  predicting the edge changes in the reactants, our  model can naturally leverage the atom embedding $h^{r}$ to  predict  an atom's other properties such as its electron charges, aromatic property and chirality.  

To this end, we  pass  the atom  embedding $h^{r}_{i}$ through a MLP layer to generate the predicted probabilities of this atom's aromatic property. 
Such prediction thus can  help us to recover the aromatic bonds, which are typically described as alternating single and double bonds, in the resulting product graphs. 

\section{Experimental Studies}
    
    
    \begin{table*}
      \caption{Top-k accuracy on USPTO-MIT; Best results in \textbf{bold}. We also show if the comparison model can be parallelly  trained in an  end-to-end fashion.  
      $^\dagger$ indicates that the results were copied from its published paper.
      The bracket indicates the method's  learning taxonomy: ``combinatorial'' for    parallel optimization,  ``graph'' for  graph translation, and ``sequence''  for an auto-regressive generation. 
      }
      \label{topk}
      \centering
      \begin{tabular}{lllllcc}
        \toprule
        \multicolumn{4}{r}{Accuracies(\%)}  \\
        \cmidrule(r){2-5}
        Model Name(scheme)  & Top-1 & Top-2 & Top-3  & Top-5  & parallel & end-to-end \\
        \midrule
        WLDN$^\dagger$ (combinatorial) & 79.6 & - & 87.7 & 89.2 & $\checkmark$  & $\times$ \\
        GTPN $^{\dagger}$( graph) & 83.2 & - & 86.0 & 86.5  & $\times$  & $\checkmark$ \\
        Transformer-base $^{\dagger}$ ( sequence)& 88.8  & 92.6 & 93.7 & 94.4 & $\times$   & $\checkmark$ \\
        MEGAN$^{\dagger}$(graph) &89.3& 92.7 &94.4 & \textbf{95.6}  & $\times$  & $\checkmark$ \\
        Transformer-augmented $^{\dagger}$( sequence)& 90.4  & \textbf{93.7} & \textbf{94.6} & 95.3  & $\times$  & $\checkmark$ \\
        Symbolic$^{\dagger}$ (combinatorial) & 90.4 & 93.2 & 94.1 & 95.0 & $\checkmark$ & $\times$\\
        \hline
        NERF  
        & \bf{90.7$\bf{\pm{0.03}}$} & 92.3$\pm{0.22}$ & 93.3$\pm{0.15}$ & 93.7$\pm{0.17}$ & $\checkmark$  & $\checkmark$ \\
        \hline
        
      \end{tabular}
    \end{table*}    
           \begin{table*}[h]
      \caption{
      Computation  speedup  (compared with Transformer) }\label{speed}
      \centering
      \begin{tabular}{llcc}
        \toprule
        Model Name  &  Wall-time &Latency & Speedup  
        \\ \hline
        Transformer (b=5) & 9min & 448ms &1 $\times$\\
        MEGAN (b=10) & 31.5min & 144ms & 0.29 $\times$ \\
        Symbolic &  $>$7h & 1130ms & 0.02 $\times$ \\
        \hline
        NERF & 20s &17ms & 27$\times$\\
        \bottomrule
      \end{tabular}
    \end{table*}
    
\subsection{Experiment Setup}
    \paragraph{Dataset}
    Most of the publicly available reaction datasets are  derived from the patent mining work of Lowe~\cite{Lowe2012ExtractionOC}, where the chemical reactions are described using a text-based representation called SMILES and RDKit library  \cite{landrum16rdkit} is used to transform the SMILES strings into molecule graphs. 
       We evaluate our model using 
       the popular benchmarking USPTO-MIT dataset. 
    This dataset was  created by~\citet{WLN2017jin} by removing duplicate and erroneous reactions from \citet{Lowe2012ExtractionOC}'s original data and filtering to those with contiguous reaction centers. The resulting dataset has about 480K samples and has been  widely used for benchmarking~\cite{schwaller2018found,WLN2017jin,do2019gtpn,bradshaw2018electro,schwaller2019moltransformer}. 
    
       
              \textbf{Data Prepossessing} For the USPTO-MIT dataset, we empirically observe that  the change of binary edges does not exceed  4. We, therefore, reduce the number of  PointerNets deployed, as described in Equation~\ref{ppointer}, from 8 to 4. That is, in this implementation, we only use 4 BondFormation and 4 BondBreaking PointerNets. 
              In addition, 
              we observe that although the computational cost is insensitive to the number of PointerNets, increasing the number of PointerNets from 4 to 8 had no accuracy benefit; on the other hand, performance degraded while decreasing the number of PointerNets  below 4.
              Also, for this dataset, there are 0.3\% of reactions do not satisfy this implementation's settings. Hence, we exclude these reactants from both training and testing, and then just subtract our predictve accuracy on the remaining reactants by  0.3\% as our model's final accuracy.
    For the node features, we follow literature for the construction. Additionally, we  include the formal charge and aromatic bond property  as part of our atom feature. 
Also, we add an extra atom feature distinguishing   reactant from reagent. We do so by  following the filtering process as in~\citep{schwaller2019moltransformer}. 

    \textbf{Model Configuration}
        We implement NERF  using Pytorch \cite{paszke19pytorch}. Both the Transformer-encoder and Transformer-decoder  contain 6 self-attention layers with 8 attention heads as in the original Transformer  configuration. 
        The node embedding dimension is 256 and the dimension of the latent representation is 64. The model is optimized with AdamW \cite{kingma14adam} optimizer at learning rate $10^{-4}$  with linear warmup and linear learning rate decay. We train the models  for 100 epochs with a batch size of 128 using two Nvidia V100 GPUs (took about 3 days).

    \textbf{Evaluation Metrics}
        Similar to \cite{WLN2017jin}, we use top-k exact  SMILES string match accuracy as our evaluation metric, which is the percentage of reactions that have the ground-truth product in the top-k predicted molecules sets. Following previous works, our experiments choose $k$ from $\{1, 2, 3, 5\}$. 
        
\textbf{Generating Multimodal Outputs}
To draw (approximate) the  top-k samples from our method  efficiently, we sample $m$ latent vectors ($m \geq k$) at increasing temperatures  and take the first $k$ different predictions as an approximation to the top-k predictions. 
Specifically, at temperature $T\in \mathbb{N}$, the latent vector is drawn from $N(\textbf{0}, T*\textbf{I})$. Using a temperature higher than 1 improves sampling efficiency by increasing diversity and therefore reducing duplicate samples. 

We use the $k$ different samples drawn at the lowest temperature for the approximation of the top-k samples since samples drawn at high temperatures tend to be noisier and less credible. 
We find this sampling  works well in practice, and provides a lower bound of the real top-k accuracy.        
        
        
        
        \textbf{Comparison Baselines} 
        We evaluate the proposed approach using the following six baselines. 
    \begin{itemize}
    \item{\textbf{WLDN}}~\cite{WLN2017jin} is a two-stage approach built upon Weisfeiler-Lehman Network, which first identifies a set of reaction centers,  enumerates all possible bond configurations, and then ranks them. 

 \item{\textbf{GTPN}}~\cite{do2019gtpn} treats a chemical reaction as a sequence of graph transformations and employs reinforcement learning to learn a policy network for such transformations.
 \item{\textbf{Transformer-base}}~\cite{schwaller2019moltransformer} leverages the power of Transformer~\cite{vaswani2017transformer} to predict SMILES strings of product graphs. 
 \item{\textbf{Transformer-augmented}}~\cite{schwaller2019moltransformer} leverages the data augmentation of SMILES strings to boost the performance of Transformer-base.
 \item{\textbf{MEGAN}}~\cite{sacha2020megan} models chemical reactions as a sequence of graph edits, and learns to predict the sequence autoregressively.
\item{\textbf{Symbolic}}~\cite{qian2020integrating} integrates symbolic inference with the help of chemical rules into neural networks.
        \end{itemize}
          We here compare with a variety of strategies  including methods deploying  parallel optimization,   graph  translation, and sequential reaction generation, as indicated by the text in the brackets in Table~\ref{topk}. 

        \subsection{Predictive Accuracy}

    Table~\ref{topk} presents the predictive accuracy obtained by the testing models  on the USPTO-MIT task.
    
    Results in Table~\ref{topk} indicate that,  
       our method outperformed all the comparison baselines in terms of top-1 accuracy, establishing a  new state-of-the-art top-1 accuracy of 90.7\% for the USPTO-MIT task.   
       As can be seen in Table~\ref{topk}, NERF  outperformed Transformer-base and MEGAN by 1.9\% and 1.4\%, respectively. Promisingly, 
   our  approach also outperformed the two-stage  optimization method with logic inference, i.e., Symbolic, and the Transformer-augmented strategy, which leverages data augmentation to boost its accuracy from 88.8\% obtained by its non-augmented version, i.e., Transformer-base. 
   

    When considering the top-2, top-3 and top-5 accuracy. The NERF 
    performed on par with 
    the state-of-the-art methods, with the worse accuracy on the Top-5 case. 
    
    We also conducted a statistical significance testing for Table~\ref{topk}.
   We ran our models with 5 random seeds. Since the variance of baseline models was not available, we conducted One Sample T-test instead of
    Paired T-test against the  Transformer model (SOTA, 90.4\%). Our T-test indicates that our model’s superior top-1 accuracy was
    statistically significant (the p-value was smaller than $10^{-5}$). 
 



\subsection{Computation Speed-up}
        
     We evaluate the computational cost of the comparison baselines with two evaluation metrics: Latency and Wall-time. 
          Latency measures 
      time needed to generate the prediction for  a single test sample. 
    The  Wall-time  pictures a more practical evaluation for modern batch-based neural models. It measures the time needed for  all the testing  samples, taking into the fact that  batch-based testing is further affected by data parallelism, e.g.,  GPU  acceleration 
    and CPU multi-core process. 
     
    We compare the Wall-time and Latency for inference with the state-of-the-art machine translation based model Molecular Transformer, graph translation model MEGAN, and rule based model Symbolic. 
    
    All models are evaluated on a single Nvidia V100 GPU and a Intel(R) Xeon(R) CPU E5-2690 v4 @ 2.60GHz with 28 cores. 
    As the  Transformer and MEGAN relies on a beam search to find the most probable prediction, the beam size $b$ is a hyper-parameter that determines the trade off between the accuracy and computational efficiency.
    We show results of  Transformer for $b = 5$, the default setting for top-5 inference, as further increasing $b$ leads to only marginal improvement in accuracy. Likewise, for MEGAN we set $b=10$. 
            All experiments uses the maximal possible batch size that fits in the GPU memory (32GB).
         We do not include the time spent on loading/saving, preprocessing and postprocessing the data. 
         More specifically, for MEGAN and Transformer, we only count the time spent on beam search. For our model, we count the time spent on computing the forward pass. For Symbolic, we count the time spent on Gurobi sampling. 
         Results are presented in Table~\ref{speed}. 
    
        Table~\ref{speed} indicates that our method achieved  at least 27 times inference speed-up when compared to the comparison  models. For example, the Transformer took 9 minutes for the Wall-time  and MEGAN and Symbolic took over 31 minutes and 7 hours respectively, while  our strategy finished the reaction generations in just 20 seconds. Also, for the Latency, all the three comparison baselines require over 100ms, yet our strategy needs only 17ms. As shown in the last  column of Table~\ref{speed}, our method achieves at least  27 times speedup over the Transformer models in terms of  Wall-time. 
        

        
            \begin{table}[h]
      \caption{ Accuracy (\%) on  individual reaction topology. Best results are in \textbf{bold}. }\label{invididualaccuracy}
      \centering
      \begin{tabular}{lllll}
       \toprule        
         & Linear & Branch  & Cyclic \\
        Sample dist. 
        & 73.7\% & 16.9\% & 0.5\%\\
        \hline
        ELECTRO & 87.0 & N/A & N/A\\
        
        Symbolic&\textbf{92.5}&83.2&68.0\\
        Transformer-base &  89.8 &  80.0 &  66.5\\
        Transformer-augmented&91.4&82.5&\textbf{74.9}\\
  \hline
        NERF & {92.2} & \textbf{85.1}& {71.4} \\     
        \bottomrule
      \end{tabular}
    \end{table}
         \begin{figure}[h]
        \centering
        \includegraphics[width = 1.0\linewidth]{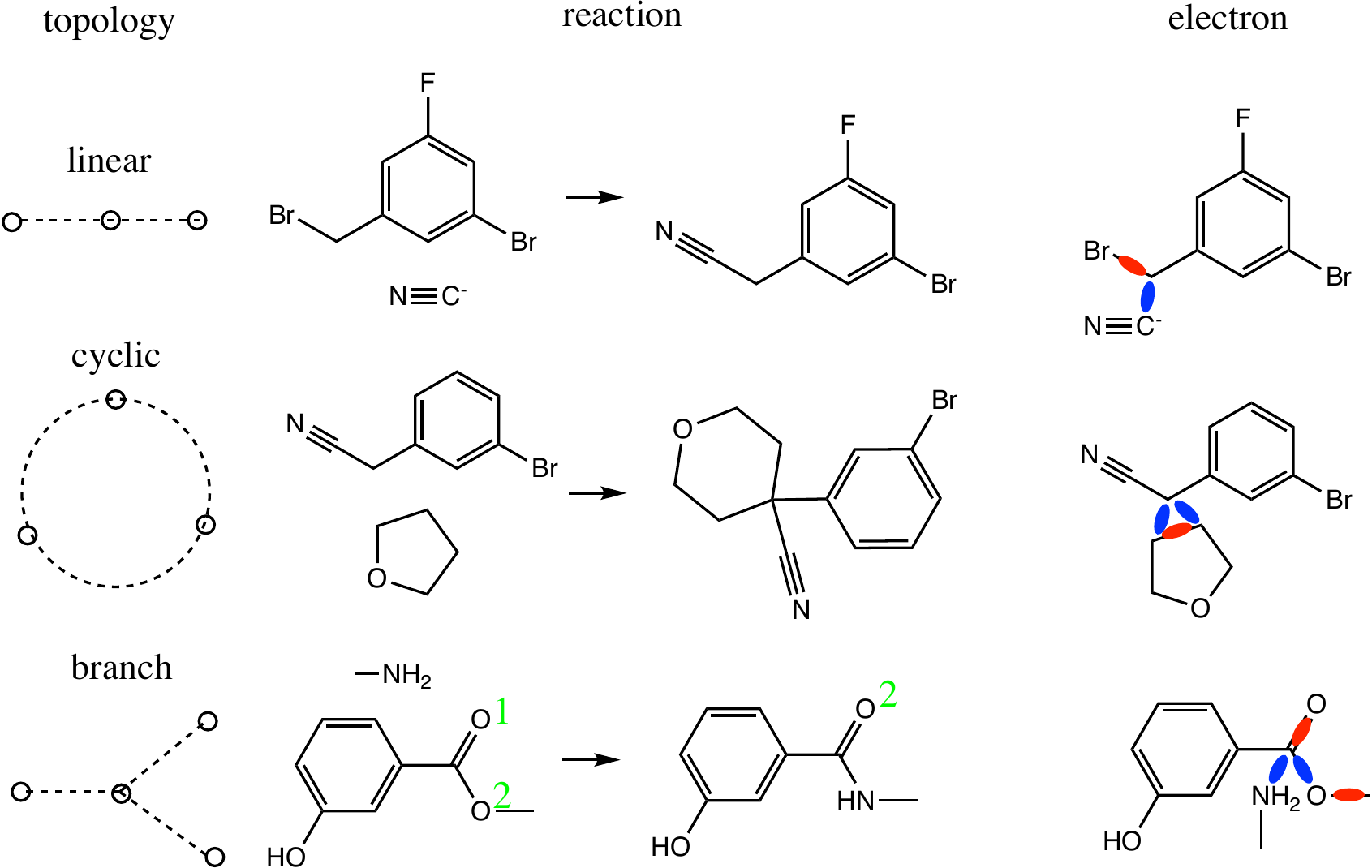}
        \caption{Different reaction topology (left) and reaction products generated by our method (middle), 
        where the green numbers distinguish different oxide atoms. 
        On the right, the blue lines denote new bonds formed, and the red lines denote  bonds broken.  
        }  
        \label{topology}
    \end{figure}
    \begin{figure}[h]
        \centering
        \includegraphics[width = 0.9\linewidth]{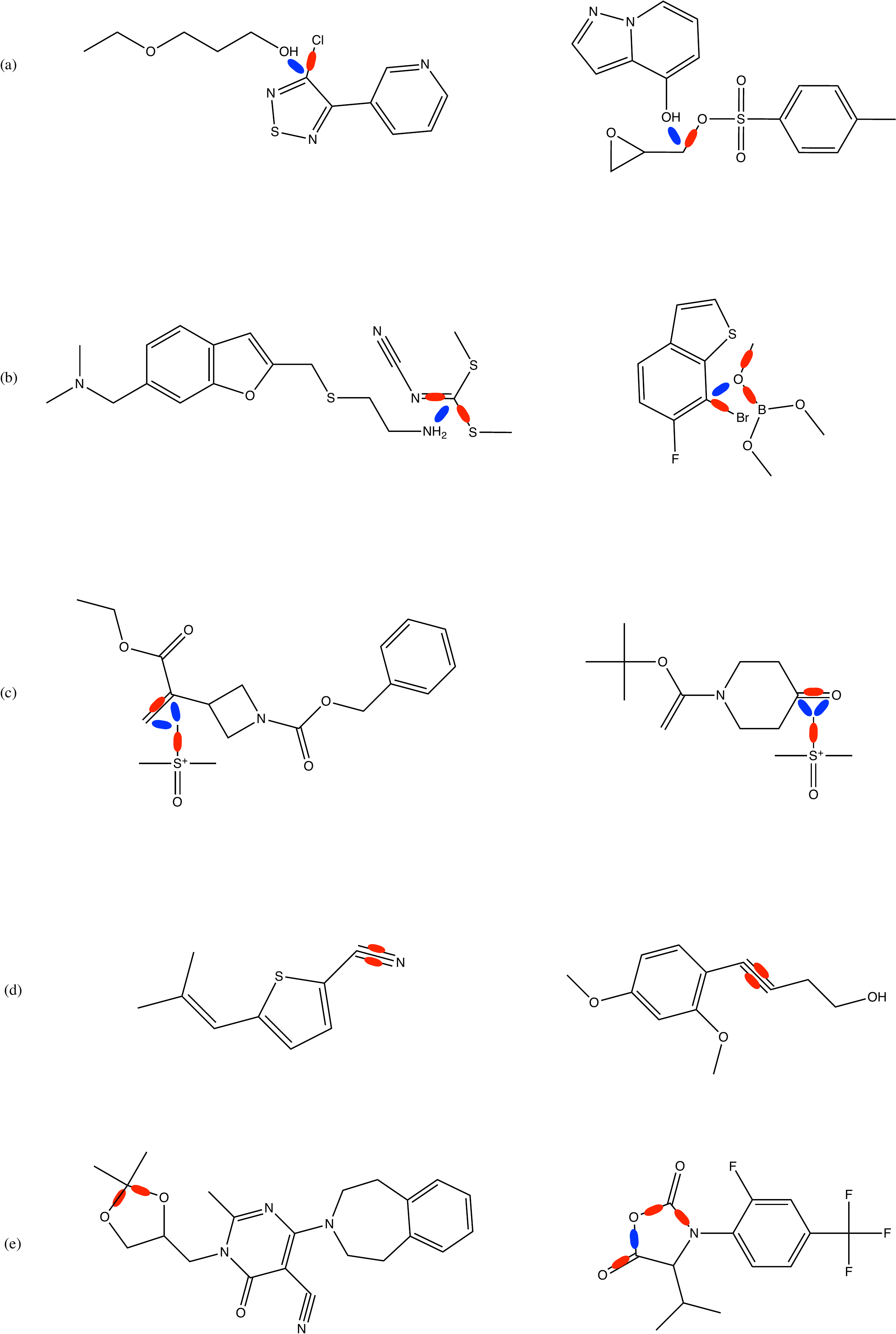}
       \caption{Extra examples of our reaction topology. (a)(b)(c)(d)(e) are samples with reaction centers~\cite{WLN2017jin} of linear, branch, cyclic, parallel and multi topology respectively. The red edges are bonds broken while blue edges denote  bonds formed.
       }
        \label{topology+}
\end{figure}
 
\subsection{Accuracy on Individual Topology}     
In addition to the overall accuracy, we also examined our model's performance on different types of reaction topology. 

        \citet{electron_flow} considers three types of   reaction topology, namely Linear, Cyclic, Complex for the full USTPO dataset  as an organizing principle for the known reactions. Nevertheless, for the USTPO-MIT dataset, this will result in extremely imbalanced categories with over 99\% of samples falling into the  Linear topology. 
        Inspired by the topology used in~\cite{bradshaw2018electro},  
        we  
         select three types of  representative topology, namely Single-Linear  (
          73\% of total samples), Branch (tree-structured, with 17\% of total samples) and Cyclic (cyclic and complex reactions with 0.5\% yet important samples) (illustrated in the left  of Figure~ \ref{topology}). 

                We compare our method with the Transformer (both base and augmented versions), Symbolic, and ELECTRO~\cite{bradshaw2018electro}. 
    Results in Table~\ref{invididualaccuracy} show that our model achieved superior or performed on par with the four comparison baselines on all the three types of topology. For example, our method achieved the best accuracy on the Linear, Branch, and Cyclic, except obtaining slightly lower accuracy than that of Symbolic on Linear and that of Transformer-augmented on Cyclic. On the other hand, the ELECTRO performed the worst on the Linear topology, while  Symbolic achieved the best on Linear type but lower accuracy on the Branch and Cyclic. Also, although Transformer-augmented obtained the highest accuracy on the Cyclic, but it obtained lower accuracy than our method on the other two categories, namely Linear and Branch. 
These results suggest that our method performs well over all the individual topology types.     

\subsection{Topology and Prediction Visualization}
 Figure~\ref{topology}  visualizes some generated reaction products that follow   the three types of topology as described above. 
The left column of Figure~\ref{topology} depicts the three types of topology, namely Linear, Cyclic, and Branch. The middle two columns show the reaction of transforming reactant on the left to the product on the right, which is correctly predicted by our method. The right  column of the figure presents the corresponding electron flows captured by our model during the reactions, where the blue lines denote new bonds formed and the red lines denote bond broken.  

Figure~\ref{topology} indicates that our model can not only predict reactions from various types of topology, but also clearly show the electron flows that result in the final reaction products.

In Figure~\ref{topology+}, we also depict additional cases of our model predictions. In addition to the aforementioned three topology, we here also list parallel and multi topology, covering all of our taxonomy. Parallel topology refers to the reactions where there are parallel edges breaking or forming, and Multi are those with multiple reaction centers, each representing a series of bond forming and breaking.

\subsection{Interpretability}    

\citet{bradshaw2018electro} show intuitive interpretability of their model due to the prediction of mechanism. 
Since our model embraces predicting  the pseudo-mechanism, namely the  electron flows during the reactions, our predictions are  also easy for chemists to interpret. 

       \begin{figure}[h]
        \centering
        \includegraphics[width = 0.9\linewidth]{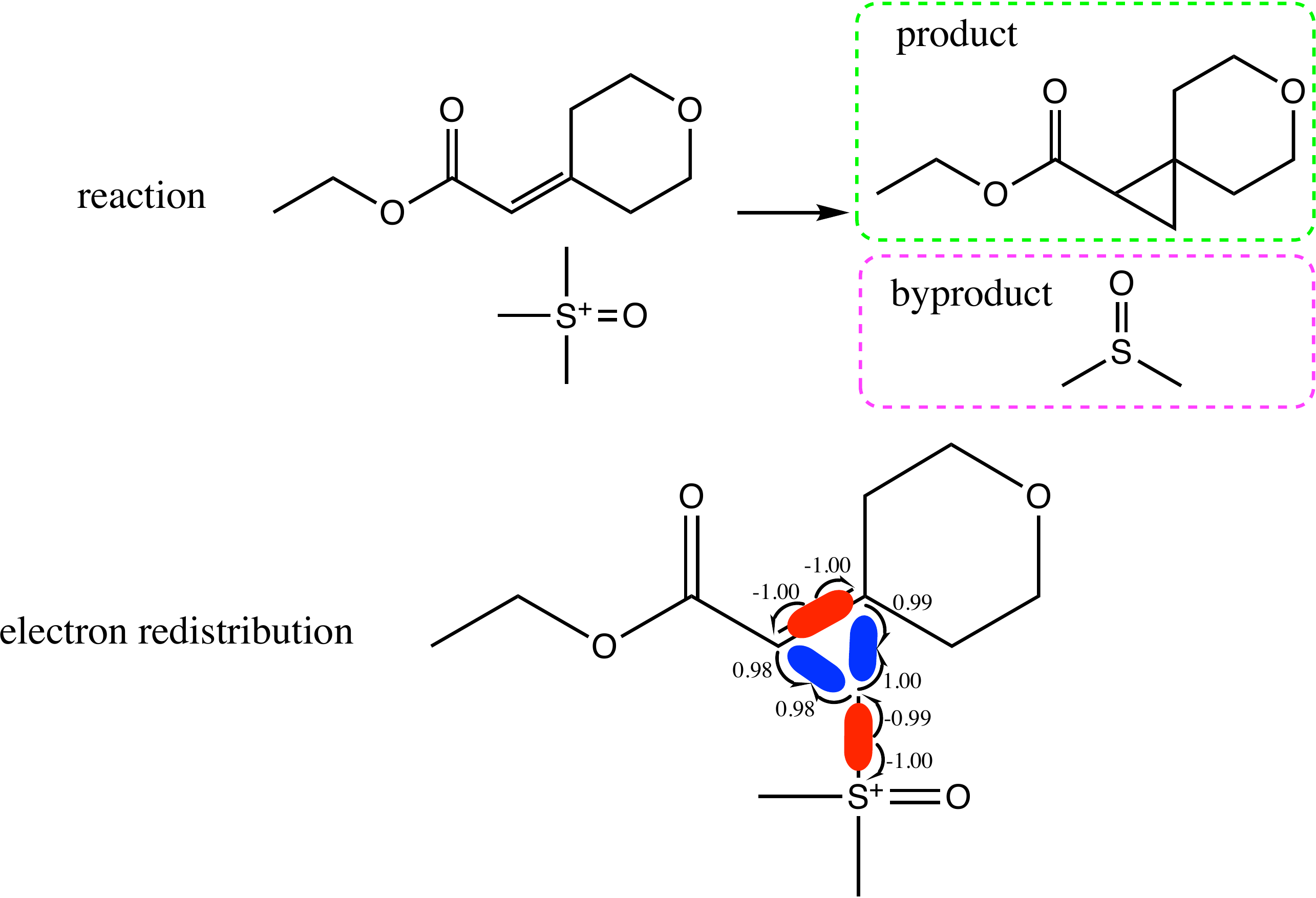}
        \caption{A reaction  correctly predicted by our model. Our model predicts the direction and the number of electron flows, which are denoted by arrow pushing and real values  respectively. 
        These predictions clearly show what caused the  bond making and breaking during the reaction, thus it is much easier for chemists to interpret.
        The product in the green box is the ground truth, which is predicted correctly, and we also predicted the  byproducts (in the purple box) at the same time when generating  the whole predicted graph $\hat G^p$.
        }  
        \label{explainable}
\end{figure}   

Figure~\ref{explainable} shows 
a reaction with non-linear topology  predicted by our model, where  the green box is the ground truth, which is predicted by our model correctly. Note that, our method  also predicted the  byproducts (shown in purple box) at the same time when generating the  whole predicted graph $\hat G^p$. 

 Figure~\ref{explainable} shows that our model correctly predicted the bond formation (in blue) and bond breaking (in red). This is explained by the predicted number   next to the pushing array,   indicating the predicted edge change $\hat e_{ij}$ as discussed in Equation~\ref{wwij}. Recall from Section~\ref{decodingsec} that by rounding $\hat e_{ij}$ to discrete values $\hat e^p_{ij}$, as described in Equation \ref{edgeGeneration}, we obtain the bond changes. 
 
 
It is worth noting that our model predicts the electron re-distribution around all atoms in the reactants. Hence, it infers side products. 
Note that the byproduct drawn as dimethylsulfoxide (DMSO) does not appear in the ground truth record for this reaction, as the dataset only contains ``major'' reaction products. Byproducts are all speculative and predicted without any labeled training examples. 
To this end, following literature, we do not evaluate the side products. 

\section{Conclusion and Future Work}

We formulated chemical reaction as a  simultaneous electron redistribution problem and  solved it with a novel  multi-pointer decoding  network. 
This results in the first  non-autoregressive electron flow model   for reaction prediction, which captures 
the simultaneous  bond making and breaking in molecules in one shot. 
We empirically verified that our method achieved  superior
top-1 accuracy  and  
  at least 27  times inference speedup  over  the state-of-the-art methods. 
 
Possessing superior predictive performance,    parallel computation, and  intuitive interpretability    makes our strategy  NERF
an appealing solution to practical learning problems involving chemical reactions. 

In the future, we will investigate the potential of  
applying our framework to more complicated settings such as stereochemistry. 


\section{Acknowledgements}
This project was supported by the Natural Sciences and Engineering Research Council (NSERC) Discovery Grant, the
Canada CIFAR AI Chair Program, collaboration grants between Microsoft Research and Mila, Samsung Electronics
Co., Ldt., Amazon Faculty Research Award, Tencent AI
Lab Rhino-Bird Gift Fund and a NRC Collaborative R\&D
Project (AI4D-CORE-06). This project was also partially
funded by IVADO Fundamental Research Project grant PRF2019-3583139727.

\bibliographystyle{icml2021}
\bibliography{preprint}
\end{document}